\documentclass[aps,pra,twocolumn,showpacs,groupedaddress]{revtex4-1}

\usepackage{graphicx,subfigure,color,verbatim,amsmath}

\renewcommand{\vec}[1]{{\bf #1}}
\newcommand{\sub}[1]{_{\mbox{\scriptsize #1}}}
\newcommand{\acf}{a}
\newcommand{\bcf}{b}

\begin{document}

\title{Vacuum Pressure Measurements using a Magneto-Optical Trap}

\author{T. Arpornthip}
\author{C. A. Sackett}
\email[]{sackett@virginia.edu}
\affiliation{Department of Physics, University of Virginia, Charlottesville, VA 22904}
\author{K. J. Hughes}
\affiliation{Triad Technology Inc., Longmont, CO 80501}

\date{\today}

\begin{abstract}
The loading dynamics of an alkali-atom 
magneto-optical trap can be used
as a reliable measure of vacuum pressure, with loading time
$\tau$ indicating a pressure less than or equal to 
($2\times 10^{-8}$~Torr s)/$\tau$.
This relation is accurate to approximately a factor of two over wide variations
in trap parameters, background gas composition, or trapped alkali species.
The low-pressure limit of the method does depend on the trap
parameters, but typically extends to the $10^{-10}$ Torr range.

\end{abstract}


\maketitle

\section{Introduction}

The use of ultra-high vacuum (UHV) systems is ubiquitous in modern atomic
physics. Vacuum practices and technologies sufficient to attain
pressures on the order of $10^{-9}$ Torr or lower are well known 
\cite{Lafferty98}, 
but nonetheless often comprise a significant experimental complexity.
In addition, the space and power requirements of UHV systems can be 
considered a barrier to the development of commercial applications based
on atomic physics techniques. While the primary concern in such systems
is the generation and maintenance of UHV pressures, 
an important secondary issue is the
measurement of pressure. 

The standard instrument for UHV pressure measurement is the 
ionization gauge \cite{Lafferty98}, 
which takes various forms and can measure pressures
to $10^{-11}$ Torr or lower \cite{Watanabe99}. 
However, ionization gauges require typically 100~W
of electrical power, and take up volumes of 100~cm$^3$ or more.
These requirements may be negligible in 
large laboratory-based vacuum systems, but as
systems
are miniaturized and streamlined to improve simplicity and efficiency,
ionization gauges are likely to become unacceptable. Another measurement
instrument, the residual gas analyzer, suffers from similar constraints.

An alternative technique is to measure vacuum pressure using
an ion pump \cite{Lafferty98, Welch01}. 
Ion pumps are primarily used to maintain vacuum pressure, but
measurement of the pump current provides a pressure indicator as well. 
Ion pumps
do not generally perform as well as ionization gauges, since
leakage currents limit the minimum pressure reading, typically in the
$10^{-9}$ Torr range. 
The relation between current and pressure is also complicated and varies 
with pump design. 
Finally, ion pumps are themselves
typically large and power intensive, and they
require a significant magnetic field near the pump, all of which
can be drawbacks in some applications \cite{DARPA}.
Pumping
methods such as evaporable and non-evaporable getters,
turbopumps, and cryopumps could avoid such problems or 
be preferred for other reasons.
None of these techniques provides a pressure measurement facility.

It is the purpose of this paper to explore the extent to which the
experiment itself can provide a pressure measurement. In particular,
we consider the magneto-optical trap (MOT), which is the starting
point for many atomic physics experiments and applications. We find
that measurement of the MOT loading time can serve as a useful and reasonably
accurate pressure gauge, but that lifetime limits imposed by
collisions between the trapped atoms give a low pressure floor in the
$10^{-10}$ Torr range.  The technique has
resolution comparable to an ion pump measurement, but avoids the
drawbacks mentioned above.
The method can be used for both beam-loaded and vapor-cell loaded traps.

We note at the outset that we do not strive here for a high-accuracy pressure
measurement. In some cases high accuracy is important, but 
a significant calibration uncertainty is usually acceptable for
vacuum diagnostic purposes. 
For instance, the sensitivity
of an ionization gauge varies by a factor of two between H$_2$
and N$_2$ gases, and by a factor of eight between He and Ar 
\cite{Lafferty98}. 
Ion pump sensitivities show similar or
greater variation \cite{Welch01}.
These effects lead to significant uncertainty if the residual gas
composition is not known. Nonetheless, both ionization gauges and ion pumps
have proven
satisfactory for many applications. We show here that MOT loading
measurements can provide an accuracy of about a factor of two, 
comparable to that typically obtained with conventional techniques.

The basic connection between MOT dynamics and background gas pressure has
been understood since MOTs were developed \cite{Raab87,Prentiss88,Bjorkholm88}, 
but to our knowledge MOTs have not previously been
proposed for providing quantitative pressure measurements.
This can likely be explained by the common availability of standard measurement
gauges. In addition, it is evident that
the relationship between background pressure and MOT dynamics will
depend on the trap depth of the MOT. The trap depth varies considerably
with the laser parameters used and can be challenging to quantify. 
This would weigh against using the MOT as a measurement tool, since 
calibration would be difficult and uncertain.

As noted by Bjorkholm \cite{Bjorkholm88}, however, the
dependence of loading time on trap depth is in fact quite weak under
most conditions. Furthermore, the loading time depends only weakly
on the type of atom being trapped and the composition of the background gas.
In light of this, a ``universal'' pressure calibration is in fact possible
so long as high accuracy is not required. Thus MOT measurements 
can serve as a convenient general-purpose measurement tool.

A similar technique was previously used by
Willems and Libbrecht \cite{Willems95} to relate the loss rate from
a magnetic trap to the pressure in a cryogenic vacuum system.
Also, collisional loss rates at a known background gas density and
MOT trap depth have been used to characterize collision cross sections
\cite{Cable90,Matherson08,Fagnan09}, or alternatively
collision rates at a known cross section and gas density can be used 
to characterize the MOT trap depth \cite{VanDongen11}.

\section{Theory}

The dynamics of MOT loading and loss are governed, to 
a good approximation, by the rate equation
	\begin{equation}
		\label{eq:PopDE}
		\frac{dN}{dt} = R-\gamma N(t) - \beta \bar{n} N(t) .
	\end{equation}
Here $N$ is the number of atoms in the trap and
$R$ is the rate at which atoms are loaded via laser cooling. Normally,
$R$ will be
proportional to the background gas pressure of the species being trapped.
The trap losses are described by 
$\gamma$, the rate constant for loss due to collisions with all
background gases, and $\beta$, the rate constant for
loss due to inelastic two-body collisions within the trap.
The two-body rate also depends on the mean density of the trapped
atoms,
$\bar{n} = (1/N)\int n(\vec{r})^2 d^3 r$.

\begin{figure}
\includegraphics[width=3.5in]{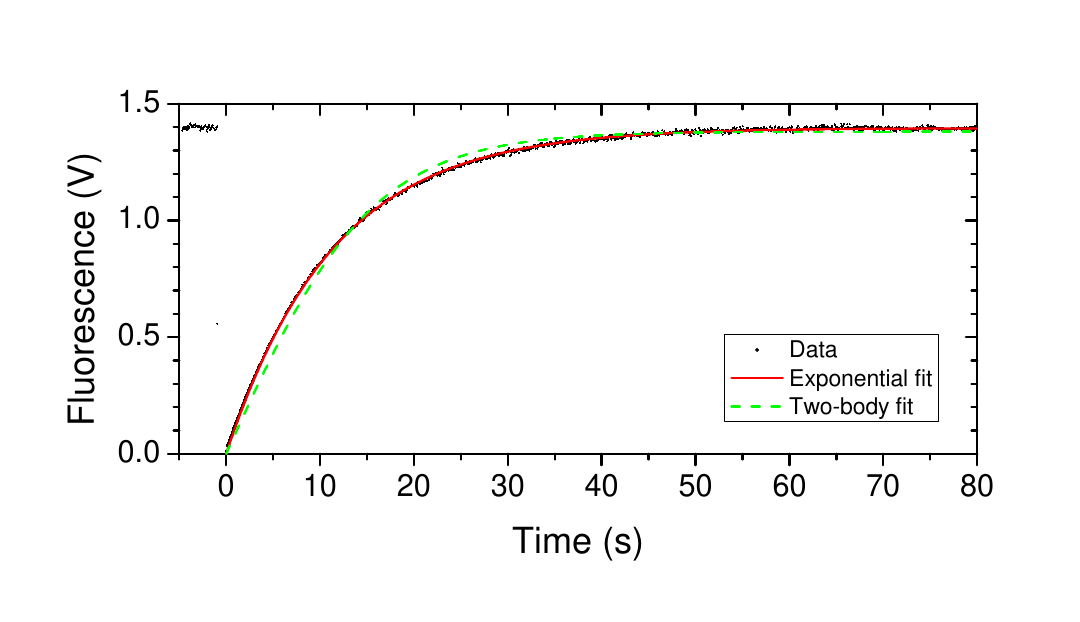}
\caption{\label{fig:LoadingCurve} 
(color online)
MOT loading dynamics. The vertical axis shows the atomic fluorescence
as measured by a photodiode, after the cooling beams are briefly blocked
at time $t = 0$. Data points are the experimentally measured values
for a $^{87}$Rb trap holding a maximum of $3\times 10^8$ atoms.
The solid (red) curve that lies on top of the data is a fit to the
exponential form of \protect\eqref{eq:expSolution}. For comparison,
the dashed (green) curve is a fit to 
$N(t) = \sqrt{RV/\beta} \tanh(t\sqrt{R\beta/V})$, the solution
to \protect\eqref{eq:PopDE} in the limit of constant volume $V$ and 
$\gamma \rightarrow 0$.
}
\end{figure}

In order to solve \eqref{eq:PopDE}, the variation of
$\bar{n}$ with $N$ must be known, which can be complicated in general.
Typically two regimes are identified, depending on the significance
of multiple-scattering forces within the MOT 
\cite{Sesko91,Townsend95,Overstreet05}.
For small $N$, less than of order $10^5$ atoms, the scattering 
forces are weak and
$\bar{n} \approx N(t)/V$ with fixed trap volume $V$. 
For larger $N$, light scattering enforces a constant $\bar{n}$
with $V \propto N$. In the constant density limit, \eqref{eq:PopDE}
results in an exponential loading curve 
	\begin{equation}
		\label{eq:expSolution}
		N = \frac{R}{\Gamma}(1-e^{-\Gamma t})
	\end{equation}
with 
\begin{equation}
\label{eq:Gamma1}
\Gamma = \gamma + \beta \bar{n}.
\end{equation}
An exponential curve is also observed in the constant volume
regime if $\gamma \gg \beta\bar{n}$, which is often the case since
$N$ is small.
For nearly all the parameters we investigated, the observed
loading curves were exponential to a good approximation.
Figure \ref{fig:LoadingCurve} shows an example. 

By measuring a curve such as Fig.~\ref{fig:LoadingCurve},
$R$ and $\Gamma$ can readily be determined. To the extent that
$\beta \bar{n}$ can be controlled or neglected, this provides
knowledge of $\gamma$, which is directly related to the background
gas density and thus the pressure. The practical impact of the $\beta \bar{n}$
term on the pressure measurement will be discussed in Section IV below.

The loss coefficient $\gamma$ can generally
be expressed as
\begin{equation}
\gamma = \sum_i n_i\langle \sigma_i v_i\rangle
\end{equation}
where the sum is over gas species $i$, with density $n_i$, speed
$v_i$, and loss cross section $\sigma_i$. The angle brackets represent
an average over the thermal distribution, and we assume that the
velocity of the trapped atoms is negligible compared to $v_i$. The 
loss cross section $\sigma_i$ is given by
\begin{equation}
\sigma_i = \int_{\theta > \theta_L} 
\frac{d \sigma}{d \Omega} d\Omega
\label{losscalc}
\end{equation}
where $d \sigma/d \Omega$ is the differential scattering cross section
and $\theta_L$ is the minimum scattering angle required in 
order to give the target cold atom sufficient energy to escape the
trap. 

The long-range 
interaction potential between ground state trapped atoms and background
species $i$
can typically be
approximated with the van der Waals form $-C_i/r^6$ \cite{Margenau69}.
The trapped atoms do have some amplitude to be in an excited state,
which typically modifies the $C_i$ coefficient. For interactions between
excited atoms and background atoms of the same species
the interaction can be significantly enhanced to an $r^{-3}$ form.
We will neglect this effect for now, since our main interest will
be losses due to vacuum contamination by non-trapped species.

Typical MOT trap depths $D$ are on the order of 1~K, which is large enough
that the cross section
can be estimated classically
but small enough that 
the small angle and impulse approximations can be 
used \cite{Bjorkholm88,Bali99}.
For a van der Waals potential this leads to
\begin{equation}
\frac{d\sigma}{d\Omega} = \frac{1}{6}
\left(\frac{15\pi}{8}\frac{C_i}{m_i v_i^2}\right)^{1/3}
\theta^{-7/3}
\end{equation}
where $m_i$ is the mass of the incident species.
The critical angle $\theta_L$ is 
$\sqrt{2m_0 D}/(m_i v_i)$
for trapped atom mass $m_0$. Evaluating 
\eqref{losscalc} then gives
\begin{equation}
\sigma_i = \left(\frac{15\pi^4}{16}\right)^{1/3}
\left(\frac{m_i C_i^2}{m_0 E_i D}\right)^{1/6}
\end{equation}
for incident energy $E_i = m_i v_i^2/2$.
Finally, $\sigma_i v_i$ can be averaged over a Maxwell-Boltzmann distribution
at temperature $T$, and the density $n_i$ can be expressed in terms of the
partial pressure $P_i = n_i k_B T$. This yields the loss rate
\cite{Bjorkholm88}
\begin{equation}
\gamma_i \approx 6.8 \frac{P_i}{(k_B T)^{2/3}}
\left(\frac{C_i}{m_i}\right)^{1/3}
(D m_0)^{-1/6}.
\label{eq:loss_coeff}
\end{equation}

Thus, as claimed in Section I, $\gamma_i$ depends only 
weakly on the trap depth $D$. 
Gensemer {\em et al.}\ and Van Dongen {\em et al.}\ measured
loss rate variations consistent with this dependence for
trap depths between 0.5 and 2~K \cite{Gensemer97,VanDongen11},
a range consistent with other reported
trap depth measurements for alkali atoms
\cite{Raab87,Kawanaka93,Hoffman96,Bagnato00}.
For the purpose of pressure measurement, we see that
the loss calibration does not vary significantly with trap
parameters such as laser intensity and detuning. Note, however,
that for smaller trap depths the loss rate can vary more significantly
as the validity of the classical approximation starts to fail
\cite{Gensemer97,Bali99,VanDongen11}.

The loss coefficient depends on the background 
gas species through $C_i$ and $m_i$.
The van der Walls coefficients $C_i$ can be estimated 
using the Slater-Kirkwood formula \cite{Margenau69,Miller77}, 
\begin{equation}
C_i = \frac{3}{2} \frac{\hbar e}{(4\pi\epsilon_0)^2 m_e^{1/2}}
\frac{\alpha_0\alpha_i}{(\alpha_0/\rho_0)^{1/2}+(\alpha_i/\rho_i)^{1/2}}
\label{slater}
\end{equation}
where $m_e$ is the electron mass and
species $i$ has static electric polarizability $\alpha_i$ and 
number of valence electrons $\rho_i$. As above,
$i = 0$ refers to the trapped species.
Typically, the polarizability of a particle increases with its mass, 
so the variation in $C_i/m_i$ is reduced compared to that of $C_i$
or $m_i$ alone.
Table~\ref{Rb_table} shows the calculated loss coefficents for trapped
Rb atoms caused by various background gas species. 

We noted previously that excited atoms generally have a different 
$C_i$ coefficient. In reference to \eqref{slater}, the dc polarizability
of alkali atoms in their first excited states is typically 
2--4 times larger than that for the ground state \cite{Miller77}.
However, the ground-state polarizability of the alkalis is already very large,
so for interactions with non-alkali species, the $\alpha_0/\rho_0$
term in the denominator typically dominates $\alpha_i/\rho_i$.
The $C_i$ coefficient therefore scales approximately as $\alpha_0^{1/2}$
and the loss rate as $\alpha_0^{1/6}$. The loss rate coefficients can
therefore be expected to differ by not more than 30\% from the
ground estimates, with the caveat that resonant interactions
can be expected to give
larger deviations 
for collisions with hot atoms of the same species as those trapped.

Finally, the loss coefficients depends only weakly on the trapped atom
species itself, as seen in Table~\ref{H2_table}. 
Note, however, that 
the classical scattering approximation may be inadequate 
for lithium atoms in a shallow MOT
\cite{Bali99}, leading to a stronger dependence on 
the trap depth.


\begin{table}
\begin{tabular}{ccll}
Species & \hspace{3ex}  & $C_i$ & $\gamma_i/P$ \\ \hline
H$_2$ & & 137 a.u. \hspace{2ex} & $4.9\times10^7$ Torr$^{-1}$ s$^{-1}$\\
He & & 35 & 2.5$\times 10^7$ \\
H$_2$O & & 241 & 2.8$\times 10^7$\\
N$_2$ & & 302 & 2.6$\times 10^7$ \\
Ar & & 278 & 2.3$\times 10^7$\\
CO$_2$ & & 482 & 2.6$\times 10^7$ \\
Rb & & 4400 & 4.4$\times 10^7$
\end{tabular}
\caption{\label{Rb_table}
Estimated loss coefficients for collisions between ground-state Rb atoms and
the indicated background gas, for a 1~K trap depth and 300~K background
gas temperature. The
$C_i$ coefficients are in atomic units, calculated using
\protect\eqref{slater}.
The loss coefficients $\gamma_i/P$ are 
calculated from Eq.~\protect\eqref{eq:loss_coeff}.
}
\end{table}

\begin{table}
\begin{tabular}{ccll}
Species & \hspace{3ex}  & $C\sub{H$_2$}$ & $\gamma\sub{H$_2$}/P$ \\ \hline
Li & & 82.5 a.u. \hspace{2ex} & $6.4\times 10^7$ Torr$^{-1}$ s$^{-1}$\\
Na & & 91 & 5.3$\times 10^7$\\
K & & 130 & 5.4$\times 10^7$\\
Rb & & 140 & 4.9$\times 10^7$\\
Cs & & 170 & 4.9$\times 10^7$
\end{tabular}
\caption{\label{H2_table}
Estimated
loss rate coefficients for collisions between ground-state alkali atoms and 
hydrogen molecules, for a 1~K trap depth and 300~K background
gas temperature. The $C\sub{H$_2$}$ coefficients are
in atomic units, taken from \protect\cite{Bali99}.
The loss coefficients are 
calculated from Eq.~\protect\eqref{eq:loss_coeff}.}
\end{table}

We conclude that in most cases, the relation between loss
rate and background pressure is expected to vary
by only a factor of about two.
For pressure measurement, this
level of variation is generally acceptable and in fact rather better than that
of conventional pressure gauges. 

\section{Measurements}

Clear experimental measurements of $\partial \gamma/\partial P$ are not
common in the literature. 
Prentiss {\em et al.}\ made an early measurement 
$\gamma/P = 5\times 10^7$ Torr$^{-1}$ s$^{-1}$ in a sodium MOT 
likely dominated by H$_2$ background
gas \cite{Prentiss88}.
More recently, Fagnan {\em et al.}\ \cite{Fagnan09} 
and Van Dongen {\em et al.}\ \cite{VanDongen11} measured
the dependence of the collisional loss in a Rb MOT on the partial
pressure of Ar gas. For a trap depth of 1~K, Fagnan obtained 
$\gamma\sub{Ar}/P = 1.6\times 10^7$ Torr$^{-1}$ s$^{-1}$ while
Van Dongen obtained $2.2\times 10^7$ Torr$^{-1}$ s$^{-1}$.
Both Ar measurements were in good agreement 
with a fully quantum calculation of the loss rate.
As seen in Tables~\ref{Rb_table} and \ref{H2_table},
our classical calculation also agrees with all of these results.

To test the relationship between vacuum pressure and loss rate ourselves,
we experimentally investigated the loading dynamics in a rubidium MOT.
The majority of the experiments were performed in a vacuum chamber
consisting of a 30-cm long, 6-cm diameter cylindrical glass cell 
that is connected to a second cell by 
a 20-cm long, 1-cm diameter tube. The MOT was produced and studied in
the cylindrical cell; the second cell is designed for the 
production of Bose-Einstein
condensates. The MOT cell is mounted on a stainless steel cross, to
which is also attached a 20 L/s ion pump from Duniway Stockroom and 
a tubulated Bayard-Alpert ionization gauge. 
The unused cell is pumped
by a 20 L/s Varian ion pump. The vacuum conduction between the two cells
is estimated as 0.5 L/s, while the conduction from the MOT region to its pump
and gauge is about 20 L/s. 
The gauge was monitored using a Granville-Phillips model 330 controller.
After a vacuum bake at 300$^\circ$C, the
base pressure reading was $3\times 10^{-10}$ Torr.
Rubidium atoms were sourced from two SAES alkali dispensers,
model Rb/NF/7/25/FT10+10, wired in series and positioned about
6~cm from the MOT location.

The main cooling laser for the MOT was an amplified 
Toptica diode laser that
generates a maximum power of 230 mW divided into six independent 
MOT beams. The beams passed through the cylindrical
wall of the MOT cell. They 
were about 4 cm diameter, yielding a maximum intensity at the atoms of 
40 mW/cm$^2$. The intensity could be reduced from this level using
an acousto-optic modulator.
The diode laser was locked to a saturated absorption cell with 
a variable detuning offset. By adjusting the lock point, either
isotope of Rb could be trapped.

The repump laser for the MOT was a home-built diode laser producing
7 mW of power, incident on the cell in a single 4-cm diameter beam.
It too was locked via saturated absorption and could be adjusted to operate
for either isotope. The intensity and detuning of the repump laser
were not changed in these experiments.

The magnetic field for the MOT was produced by a pair of coils, giving
a gradient of 7~G/cm in the vertical and 3.5~G/cm in the horizontal directions.

Finally, the fluorescence from the MOT was monitored using 
a photodiode. Light is collected with a solid angle of 
$2\times 10^{-3}$~srad and
converted to a voltage with an efficiency of 2 V/$\mu$W.
The fluorescence measurements are used to estimate the atom number
via the scattering rate
\begin{equation}
R\sub{scat} = \frac{\Gamma_a \Omega^2}{2\Omega^2+4\Delta^2+\Gamma_a^2}
\end{equation}
for atomic linewidth $\Gamma_a = 2\pi\times 6$~MHz, 
laser detuning $\Delta$,
and Rabi frequency $\Omega$ given by $\Omega^2 = \Gamma_a^2 I/I_s$ for
laser intensity $I$ and saturation intensity $I_s \approx 3.2$~mW/cm$^2$.

\begin{figure}
\includegraphics[width=3.5in]{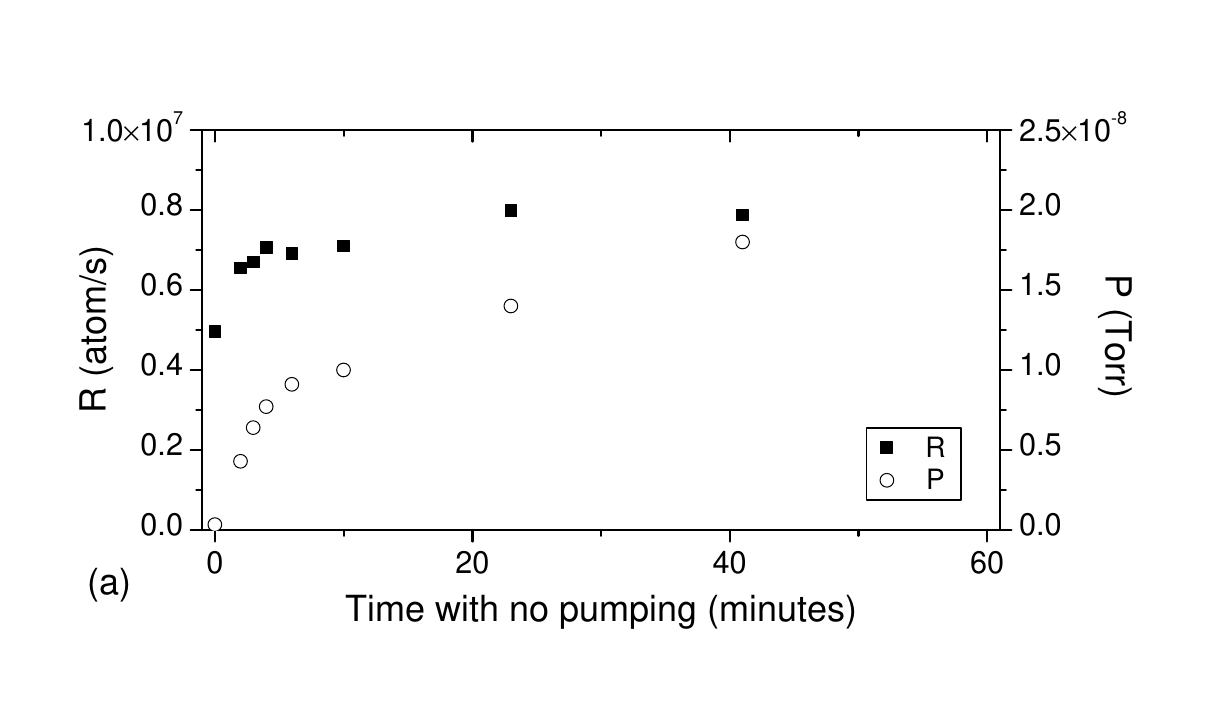}

\vspace{-0.5in}

\includegraphics[width=3.5in]{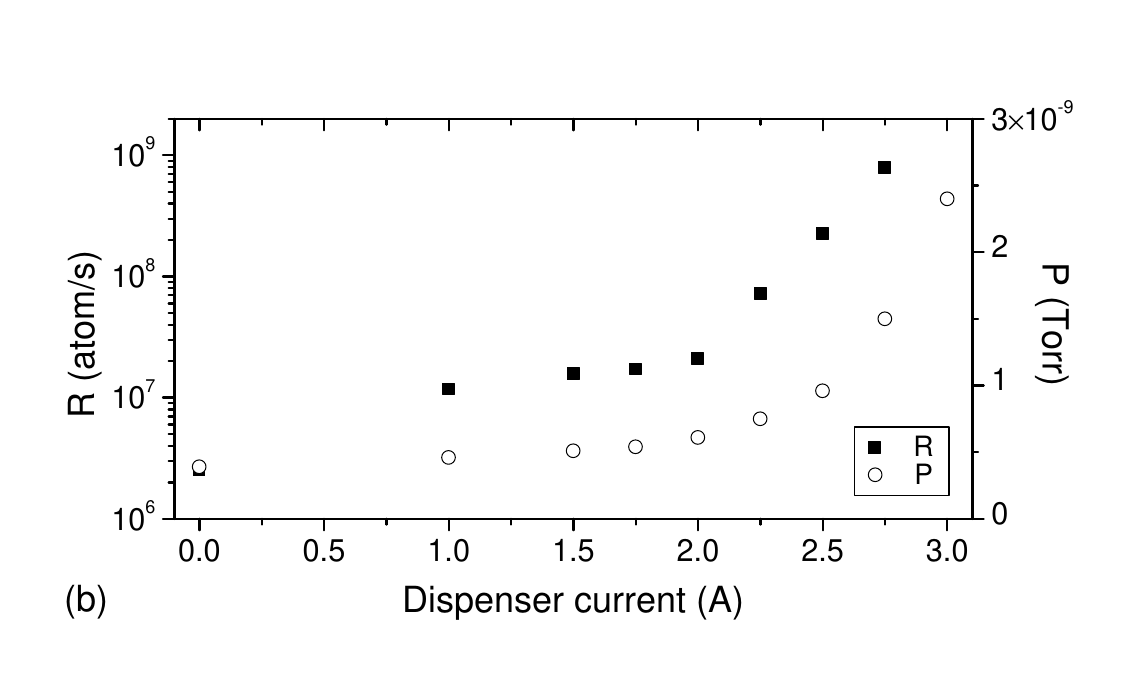}
\caption{Response of the pressure gauge reading $P$ and the MOT loading
rate $R$ to (a) turning off the chamber pumps and (b) increasing
the Rb dispenser current.
\label{pressure_rate}
}
\end{figure}

We investigated how the loading and loss rates
varied with the system pressure.
One way to vary the pressure is to turn off the 
ion pumps (in both cells).
The resulting behavior of the pressure and the MOT load rate
are shown in Fig.~\ref{pressure_rate}(a). The fact that the relative change
in $P$ is much larger than the relative change in $R$ indicates that
the partial pressure of Rb remains fairly constant while the pressure
due to other gases increases.

Another way to vary the pressure is to operate the Rb dispensers.
Figure~\ref{pressure_rate}(b) shows the response as a function of current,
after allowing the system to equilibrate for 40 minutes after each
change. Here the relative change in $P$ is small compared to the
change in $R$. This is perhaps surprising, 
since the ionization gauge sensitivity
to Rb is quite high, about twenty times larger than the sensitivity to H$_2$
\cite{Summers66}. 
Also, both the measurements of MOT loss rates described below and 
the observation of Rb vapor fluorescence indicate that when the 
dispenser is running at higher currents, Rb is the dominant gas species present.
It might therefore be expected that the gauge reading would scale
with the loading rate.

We interpret the lack of such scaling to mean that Rb atoms are predominantly 
gettered by the chamber walls before reaching the gauge.
At low surface coverage, the binding energy between
alkali atoms and metal substrates is of order 3 eV \cite{Aruga89},
and binding energies to glass are expected to be similar
\cite{Garofalini88}.
The vapor pressure resulting from such bonds will be negligible
at room temperature. Rubidium will also react chemically with
water and other surface contaminants.
It is thus plausible that the probability for
Rb atoms to make their way from the MOT region to the
gauge is relatively low. 

Of course,
the gettering effect will become saturated as Rb coverage builds up.
For the dispenser emission rates used here, it would require hours or days
to deposit one monolayer of Rb over the entire surface area of the MOT 
chamber. Because we operate the dispensers only for a few hours per day 
on average,
the ion pump can be expected to maintain the chamber surfaces in a mostly clean
state where gettering is effective.

We tested the surface gettering interpretation by
running a dispenser that was attached to a pumping station with
a residual gas analyzer. Conductance from the dispenser to the analyzer was
about 0.3 L/s and from the analyzer to the pump was 15 L/s.
The dispenser was run until Rb metal was observably deposited on 
a glass surface, indicating a partial pressure comparable to the vapor 
pressure of bulk Rb, $4\times 10^{-7}$ Torr. 
No Rb peaks were observed at the analyzer, with a sensitivity
of $1\times 10^{-13}$ Torr. 
This indicates an effective pumping speed for Rb of at 
least $10^6$~L/s, which would seem to require pumping action by the chamber 
walls.

This explanation implies that the total vacuum pressures
measured at the gauge and at the MOT location can be
significantly different, raising the question of exactly which pressure
is to be determined with our technique.
We believe that,
in practice, it is the pressure coming from non-Rb species
that is of greatest interest to a vacuum system designer. 
Most 
systems will anyway provide a way to control the partial pressure of the
species being studied, meaning that the loading and loss rates for 
that species can be optimized for the application at hand
whether a pressure gauge is available or not. 
A gauge is instead typically used for diagnosing problems
such as vacuum leaks, contaminated surfaces, or 
insufficient pump capacity, all of which impact the background
gas pressure. We therefore focus on the pressure $P$
as measured by the ionization gauge, and treat $P$ and $R$ as effectively
independent variables in our analysis. (This now justifies our neglect
in Section II of excited state collisions between identical atoms.)

\begin{figure}
\includegraphics[width=3.5in]{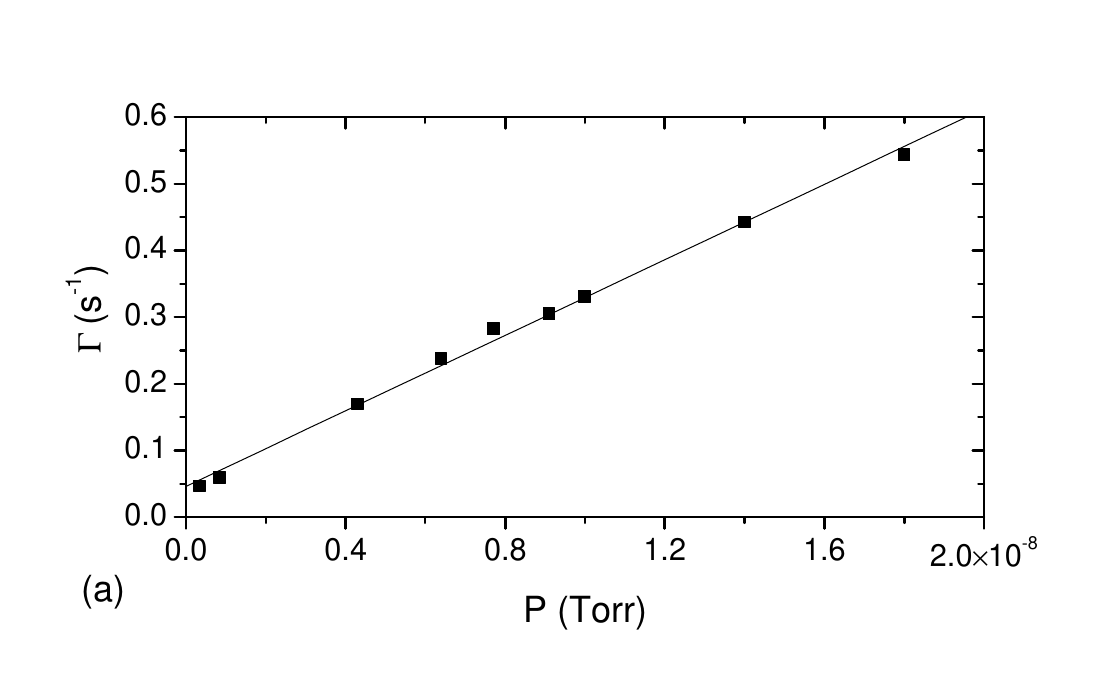}

\vspace{-0.5in}

\includegraphics[width=3.5in]{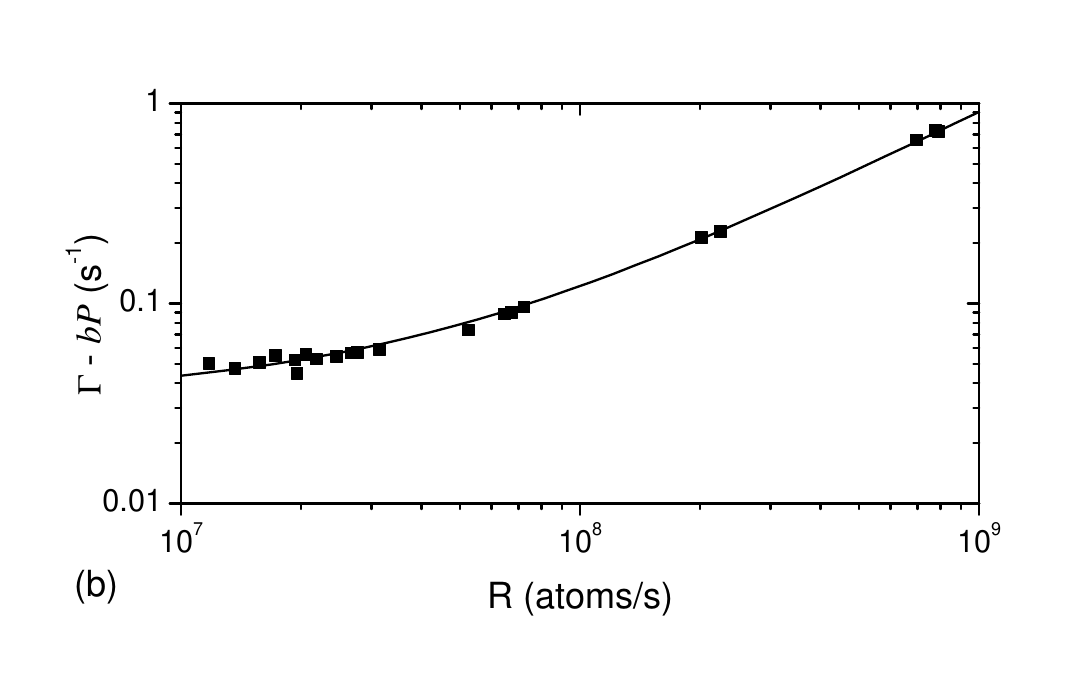}

\vspace{-0.5in}

\includegraphics[width=3.5in]{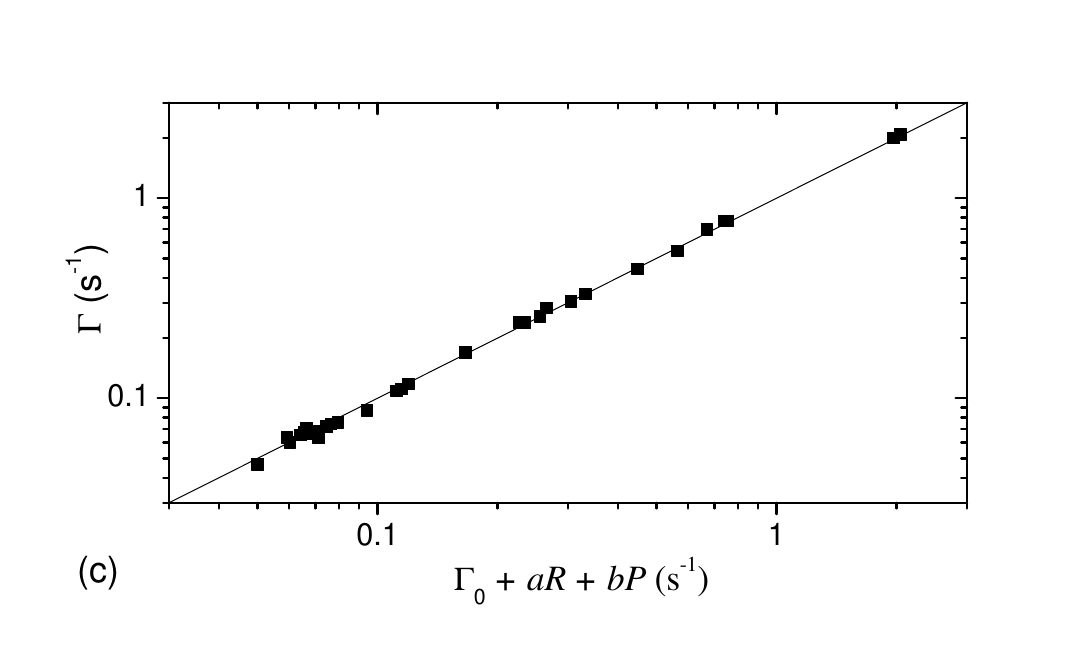}
\caption{Dependence of the MOT loss rate $\Gamma$ on the load rate $R$
and gauge pressure $P$. (a) Response to turning off the chamber pumps
as in Fig.~\protect\ref{pressure_rate}(a). The line is a linear fit.
(b) Response
to increasing the dispenser current as in Fig.~\protect\ref{pressure_rate}(b).
The curve is a linear fit.
(c) Comparison
between the measured loss rate $\Gamma$ and the model loss rate
$\Gamma\sub{fit} = \Gamma_0 + \acf R + \bcf P$, for fit parameters 
$\Gamma_0 = 0.036$~s$^{-1}$, $\acf = 8.55\times 10^{-10}$, and
$\bcf = 2.9\times 10^7$~Torr$^{-1}$s$^{-1}$. The line shows
$\Gamma = \Gamma\sub{fit}$.
\label{Gamma_PR}
}
\end{figure}

Figure \ref{Gamma_PR}(a) shows how the MOT loss rate $\Gamma$
varied with pressure after our ion pumps were turned off as in
Fig.~\ref{pressure_rate}(a). 
The clearly linear relationship can be described by 
$\Gamma = \Gamma_0' + \bcf P$, as expected.
When the dispensers are activated, the loss rate increases further,
as seen in Fig.~\ref{Gamma_PR}(b). The solid line is a linear fit
to the form $\Gamma - \bcf P = \Gamma_0 + \acf R$, 
indicating a relationship
\begin{equation}
\Gamma = \Gamma_0 + \acf R + \bcf P
\end{equation}
as seen in Fig.~\ref{Gamma_PR}(c). 
In relation to \eqref{eq:Gamma1}, here $\Gamma_0$ accounts for two-body
losses $\beta \bar{n}$, $\bcf$ accounts for collisional losses
due to background gases, and $\acf$ accounts for collisional losses
due to hot Rb atoms. If $\bar{n}$ varies with $R$,
the $\acf R$ term would also include that variation to first order.

Fitting both data sets together yields
values $\Gamma_0 = 0.036 (9)$~s$^{-1}$, 
$\acf = 8.5(1)\times 10^{-10}$, and 
$\bcf = 2.9 (2)\times 10^7$ Torr$^{-1}$ s$^{-1}$.
The error values listed
represent the fit uncertainty, estimated from the parameter
variation required to increase $\chi^2 = \sum (\Gamma-\Gamma\sub{fit})^2$
by a factor of two from its minimum value. 
In particular, we note that the $\bcf$ parameter is in
reasonable agreement with the theoretical calculation of 
Section~2.

\begin{figure}
\includegraphics[width=3.5in]{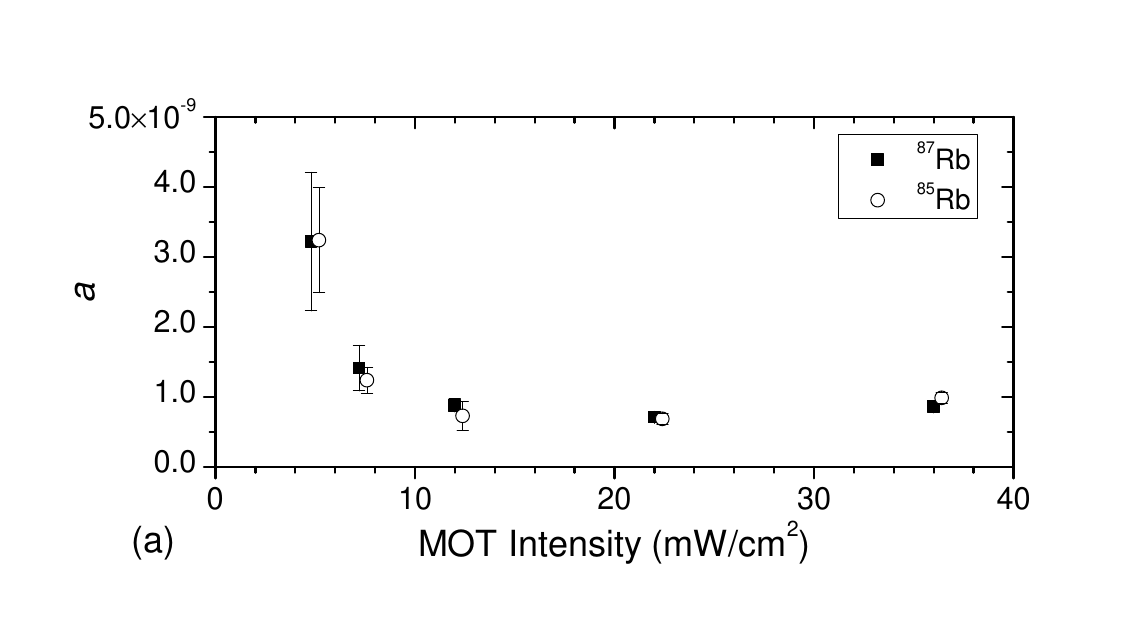}

\vspace{-0.4in}

\includegraphics[width=3.5in]{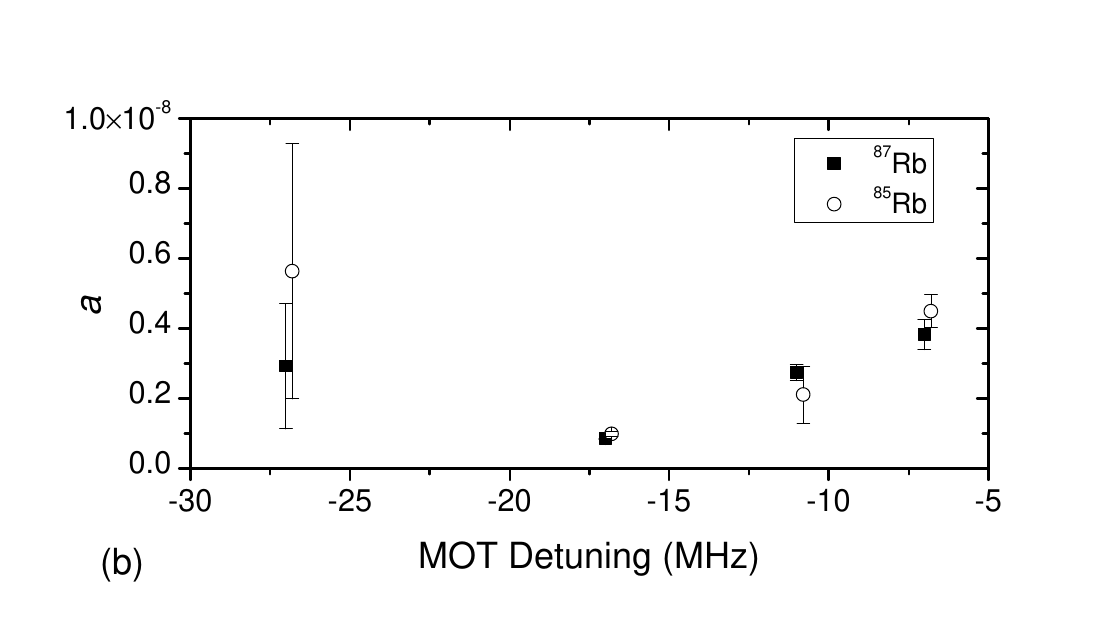}

\vspace{-0.3in}

\caption{Loss parameter $\acf = \partial\Gamma/\partial R$ as
a function of (a) total intensity and (b) detuning, for each Rb isotope.
Points for the two isotopes are offset slightly for clarity.
\label{fig:a}
}
\end{figure}

\begin{figure}
\includegraphics[width=3.5in]{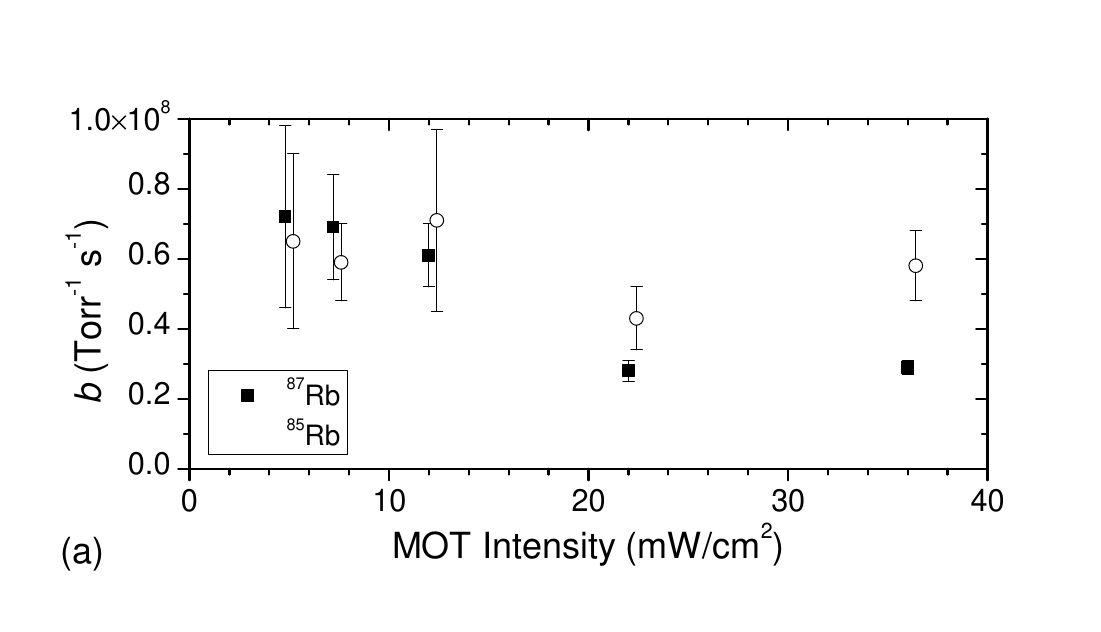}

\vspace{-0.4in}

\includegraphics[width=3.5in]{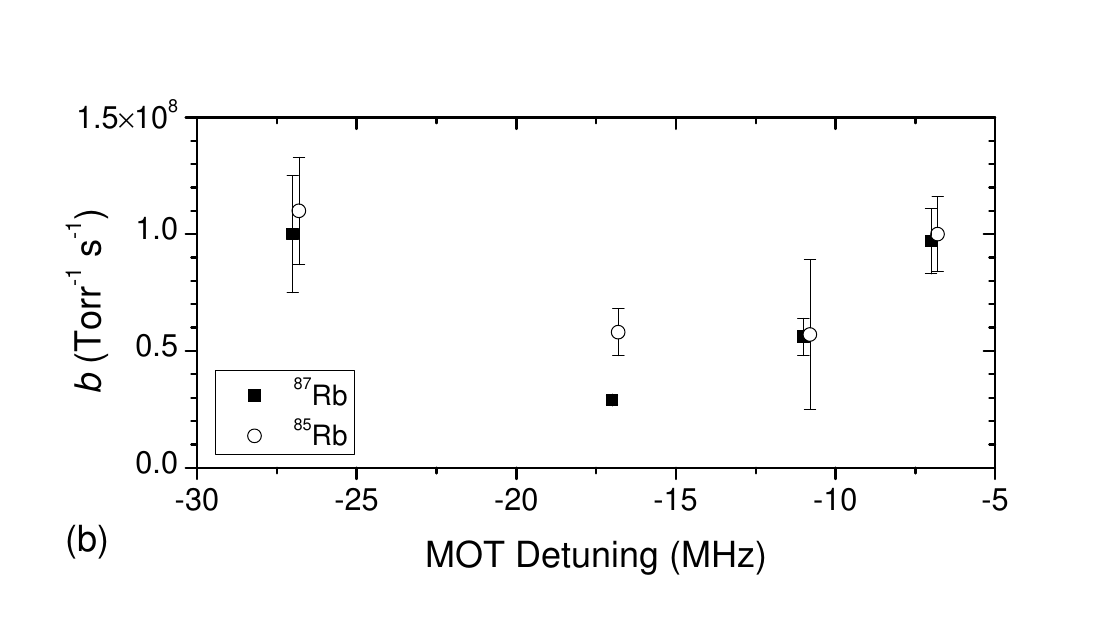}

\vspace{-0.3in}

\caption{Loss parameter $\bcf = \partial\Gamma/\partial P$ as
a function of (a) total intensity and (b) detuning, for each Rb isotope.
Points for the two isotopes are offset slightly for clarity.
\label{fig:b}
}
\end{figure}

The data of Fig.~\ref{Gamma_PR} were all taken under identical conditions
for the MOT, with an intensity of 40 mW/cm$^{2}$ and 
a detuning of \mbox{-17~MHz}.
Figures~\ref{fig:a} and \ref{fig:b} show how the loss coefficients
vary under changes in the intensity,
detuning, and choice of isotope. 
In most cases, the data
are dominated by large loss rate values and the fits give
$\Gamma_0$ consistent with zero. The observed loss rates 
at low $R$ and $P$ ranged from 0.01 to 0.1 s$^{-1}$ for $^{87}$Rb and
0.1 to 0.2 s$^{-1}$ for $^{85}$Rb.

It is reasonable that 
$\acf$ should diverge
as the laser intensity approaches zero, 
because the load rate $R$ will vanish even if 
$\gamma\sub{Rb}$ remains constant. 
Figure~\ref{fig:a}(a) exhibits this behavior, but 
otherwise the variation in $\acf$ and $\bcf$ is modest, as expected
from the analysis of Section II. In particular, the variation in 
$\bcf$ by roughly a factor of two over a large range of MOT parameters
supports its utility for pressure estimation.
An error-weighted average of all the data gives 
$\bcf = 5 (2)\times 10^7$~Torr$^{-1}$ s$^{-1}$ where the uncertainty
is taken as the standard deviation of the values. 
In terms of the loading time $\tau = 1/\Gamma$, we have
$P \approx 2 \times 10^{-8}$ Torr s/$\tau$.

We also checked the dependence of $\bcf$ on beam diameter and magnetic field
gradient. We observed no change in the trap loading time for beam diameters
as small as 1.5~cm, or for magnetic field gradients in the range of
5 to 10 G/cm.

To confirm the reliability of the loss-pressure calibration,
we performed similar measurements in two other laser cooling apparatuses.
The first featured a vacuum system similar to the one detailed above,
but with independent lasers and optics, larger laser beams, 
a Varian ion pump in place of the Duniway model,
and a different ionization gauge controller. 
Under its normal operating conditions with
$^{87}$Rb, 
this system gave $\bcf = 5 (1)\times 10^{-7}$ Torr$^{-1}$ s$^{-1}$, in
good agreement with the measurements in the original chamber.
Here the measured value of $\Gamma_0$ was 0.12~s$^{-1}$.

The second alternate system belonged to another research group
and was considerably different.
Here a $^{85}$Rb
MOT was formed in a 15-cm diameter, 10-cm wide cylindrical stainless 
steel chamber with glass windows. It was pumped by a 15 L/s Gamma Vacuum
ion pump. The pump and an ionization gauge were attached to the
chamber through a conflat cross with an estimated
conductance of 40~L/s for each. 
A single rubidium dispenser was mounted on a second, similar, cross.
This MOT used three retro-reflected cooling beams.
Under normal operating conditions, the 
system gave $\bcf = 6 (2)\times 10^{7}$ Torr$^{-1}$ s$^{-1}$, 
again in good agreement with our other results. 
At low $R$ and $P$, we found $\Gamma_0 = 0.3$~s$^{-1}$.

\section{Discussion}

The preceding theoretical and experimental observations lead to the
conclusion that measurements of MOT loading times can indeed 
provide a useful indicator of vacuum pressure. In practice,
measurements should be made at low load rates so that $\acf R$ is
negligible. The pressure sensitivity will then be limited by the
two-body loss term $\Gamma_0$: if $\bcf P$ is small compared to 
$\Gamma_0$, then the loading time will be nearly independent of the pressure.

In principle, $\Gamma_0 = \beta \bar{n}$ can be determined and subtracted from
the total loss rate to 
increase the pressure sensitivity. 
Experimental and theoretical estimates
for $\beta$ are available \cite{Gensemer97}, 
and $\bar{n}$ can be measured. Unfortunately,
accurate density measurements are difficult, and $\beta$ does
depend significantly on the MOT parameters.
Gensemer {\em et al.}\ report $\beta \approx 10^{-11}$~cm$^{3}$~s$^{-1}$
for both Rb isotopes
at an intensity of 40~mW/cm$^2$ and detuning of -17~MHz \cite{Gensemer97}. 
Estimating
our density at $10^{10}$~cm$^{-3}$ gives $\beta\bar{n} = 0.1$, 
compared to our observations of $\Gamma_0 = 0.04$~s$^{-1}$ for 
$^{87}$Rb and 0.1 for $^{85}$Rb. We also observe the $\Gamma_0$ rate
to vary significantly with beam alignment, presumably
due to variations in the density.

This issue can be circumvented in experiments with a MOT that is loaded from
a beam, or by another method that can be rapidly turned off. In this case,
the MOT can be filled, the loading turned off, 
and the subsequent decay observed.
Intra-trap collisions may 
cause the decay to be non-exponential at first, but as 
the density is reduced an exponential regime is reached where losses are
dominated by background collisions. 
If this regime can be observed, the
background loss rate can be determined directly. 
This technique was used, for instance, by Prentiss 
{\em et al.}~\cite{Prentiss88}.

In a typical vapor-loaded MOT such measurements are not possible,
so extending the pressure
sensitivity below the limit set by $\Gamma_0$ will be difficult.
Nonetheless, we observed $\Gamma_0$ as low as 0.013~s$^{-1}$ for
$^{87}$Rb using a total intensity of 4~mW/cm$^2$, a detuning of -17~MHz,
and carefully aligned and power-balanced beams. 
This corresponds to a pressure limit
of $2.5\times 10^{-10}$ Torr, which 
is less sensitive than possible with an ionization gauge, but
better than typically achieved with an ion pump.
This also corresponds to our system's base pressure, suggesting
that the loss rate is still pressure limited here.
Similar MOT loss rates
have been observed by other groups for all the alkali atoms
\cite{Raab87,Prentiss88,Sesko91,Kawanaka93,Myatt96,Williamson97,Gensemer97}, 
with background pressures (when reported)
of $2\times 10^{-10}$ Torr or lower, as expected.
The lowest reported MOT loss rate we are aware of is about 1 hour$^{-1}$ for
a cesium MOT in a cryogenic chamber \cite{Willems95}. 
This corresponds to a room temperature
pressure of $5\times 10^{-12}$ Torr, exceeding the sensitivity of 
most commercial ionization gauges. 

In practice it would be difficult to 
know whether an observed loading time was in fact limited by pressure
or by pressure-independent losses. For instance, one of the alternate
systems we measured had $\Gamma_0 = 0.3$~s$^{-1}$, 
indicating a pressure of $6\times 10^{-9}$ Torr. This substantially exceeds 
the measured pressure $1\times 10^{-9}$ Torr, 
presumably due to a large inelastic collision loss rate for the
the laser parameters and beam alignment used in that system. 
Without a pressure gauge,
however, it would only be possible to say that the pressure was
at most $6\times 10^{-9}$ Torr. Given the difficulty in predicting
$\Gamma_0$ for a given system, the absolute sensitivity of the MOT
technique is difficult to quantify. It would seem, however, that if
care is taken to adjust the MOT parameters to make $\Gamma_0$ as
small as possible, sensitivities in the $10^{-10}$ Torr range are
achievable.

At the other extreme, the technique will fail at high pressures
when it is not possible to achieve a MOT. This would be particularly
pernicious when it is not clear whether the lack of
a MOT is due to poor vacuum or to some other problem. At high
dispenser currents, we observed MOTs with $\Gamma$ up to
20~s$^{-1}$. Losses here were clearly dominated by collisions
with hot Rb atoms, but the corresponding background pressure $\Gamma/b$ is
$P = 4\times 10^{-7}$ Torr. The largest background pressure we obtained
by leaving the pumps turned off was $1\times 10^{-7}$ Torr, at which point the 
MOT still functioned.

In summary, we hope to have illustrated here that MOT loading
times can provide a reasonably reliable and accurate measurement of 
background pressure in a UHV system. The procedure
is relatively straightforward: With the MOT loading rate $R$ as small
as possible, adjust the lasers and other MOT parameters to make the
loading time $\tau = 1/\Gamma$ as large as possible. The vacuum pressure is
then at most
$(b\tau)^{-1} = (2\times 10^{-8}$ Torr s)/$\tau$, 
roughly independent of the
MOT laser parameters, background gas species, or trapped alkali species.
We expect this technique to be useful in situations where a conventional
pressure gauge is impractical due to other constraints on the
vacuum system design.

\begin{acknowledgments}

We thank T.~F.~Gallagher and Hyunwook Park
for taking and sharing data from their MOT system.
We are grateful to M.~F.~Francis, R.~A. Horne, and R.~H. Leonard
for helpful comments and discussion.
This project was supported by the Small Business Innovation Research (SBIR) 
program of the U.S. Navy under contract number N68335-10-C-0508.

\end{acknowledgments}


\end{document}